\documentclass[slac_one]{revtex4}
\usepackage{graphicx}
\usepackage{fancyhdr}
\newcommand{\ep}{\varepsilon}
\newcommand{\Li}[2]{{\mbox{Li}}_{#1}\left(#2\right)}
 
\newcommand{\SN}[3]{{\mbox{S}}_{#1,#2}\left(#3\right)}

\newcommand{\Snp}[2]{{\mbox{S}}_{#1\!}\left(#2\right)}
\begin{document}
\title{Differential Reduction Algorithms for the All-Order Epsilon Expansion of Hypergeometric Functions}
 
\author{S.A.~Yost} 
\affiliation{Department of Physics, The Citadel, 171 Moultrie St.,
 Charleston, SC 29409, USA}
\author{M.Yu.~Kalmykov}
\affiliation{~~{II.}\ Institut f\"ur Theoretische Physik, Universit\"at,
Hamburg,
Luruper Chaussee 149, 22761 Hamburg, Germany 
\hfill\break
and Joint Institute for Nuclear Research,
141980 Dubna (Moscow Region), Russia}

\author{B.F.L.~Ward}
\affiliation{Department of Physics, Baylor University,
 One Bear Place, Waco, TX 76798, USA}
 
\begin{abstract}
Hypergeometric functions provide a useful representation of Feynman
diagrams occuring in precision phenomenology. 
In dimensional regularization,
the $\ep$-expansion of these functions about $d=4$ is required. We discuss 
the current status of differential reduction algorithms. As an illustration,
we consider the construction of the all-order $\ep$-expansion
of the Appell hypergeometric 
function $F_1$ around integer values of the parameters
and present an explicit evaluation of the first few terms. 
\end{abstract}

\maketitle
\thispagestyle{fancy}
\vspace{-1em}
%%%%%%%%%%%%%%%%%%%%%%%%%%%%%%%%%%%%%%%%%%%%%%%%%%%%%%%%%%%%%%%%%%%%%%%
It is well-known that the hypergeometric representation is one of the most 
fruitful tools in the investigation of analytical properties of Feynman diagrams, and for efficiently evaluating these diagrams.  The manipulation of 
hypergeometric functions can be separated into two distinct problems: \vspace{-0.5em}
\begin{itemize}
\item[]
(A) deriving relations between hypergeometric functions with different parameters; \vspace{-0.5em}
\item[]
(B) finding an algorithm for the construction of the Laurent expansion of 
hypergeometric functions.\vspace{-0.5em}
\end{itemize}
The first problem was solved by mathematicians \cite{D}, 
wheres the second one, which has been analyzed mainly by physicists, is still 
a subject of active research.  The series of our recent papers has been devoted
to the latter problem. \cite{KWY07a,KWY07b,KWY07c}
There are three different ways to describe special functions
of the type occurring in Feynman diagrams: \vspace{-0.5em}
\begin{itemize}
\item[] (i)  as an integral of the Euler or Mellin-Barnes type;\vspace{-0.5em}
\item[] (ii) by a series whose coefficients satisfy certain recurrence relations; \vspace{-0.5em}
\item[] (iii) as a solution of a system of differential and/or difference 
equations (holonomic approach). \vspace{-0.5em}
\end{itemize}
For functions of a single variable, all of these representations are 
equivalent, but some properties of the function may be more evident
in one representation than another.  These three different representations 
have led physicists to three separate approaches to developing the 
$\ep$-expansion of hypergeometric functions in Feynman diagram calculations. 

The most impressive result in the Euler integral representation was 
the construction of the all-order 
$\ep$-expansion of Gauss hypergeometric functions with special values of 
parameters in terms of Nielsen polylogarithms \cite{hyper:partial}.  
Such Gauss hypergeometric functions are related to 
one-loop propagator-type diagrams with arbitrary 
masses and momenta, two-loop bubble diagrams with arbitrary masses, and 
one-loop massless vertex-type diagrams. 

The series representation has also been very useful and intensively 
studied.  The first results of this type were derived in context of 
the so-called ``single scale'' diagrams \cite{series}. 
Particularly impressive results involving series representations were 
derived in the framework of ``nested sums'' in Ref.\ \cite{nested} 
and in a generating function approach in Refs.\ \cite{KWY07b,KK08a}.

The differential equations satisfied by hypergeometric functions
provide another approach.
Results for Gauss hypergeometric functions expanded about
integer and half-integer values of parameters were presented in 
Ref.\ \cite{diff_red,KWY07a}, and results for a special type of 
rational coefficients in Ref.\ \cite{KK08a}, while results 
for generalized hypergeometric functions ${}_pF_{p-1}$ 
with integer values of parameters were presented in Ref.\ \cite{KWY07c}.
The iterated solution approach provides an
efficient algorithm for constructing the $\ep$-expansion, since each new 
term is related to previously known terms.

An important tool for the iterated solution is the iterated integral 
defined for each $k$ by 
$ I(a_k, a_{k-1},\ldots , a_1;z) = 
\int_0^{z} \frac{dt}{t-a_k} I(a_{k-1},\ldots , a_1;t).$ 
A special case of this integral,
\vspace{-0.5em}
\[
G_{m_k,m_{k-1},\ldots , m_1}(a_k, \ldots ,a_1;z)
\equiv
I(\underbrace{0, \ldots , 0}_{m_k-1 \mbox{ times}}, a_k, \cdots, 
\underbrace{0, \ldots , 0}_{m_1-1 \mbox{ times}}, a_1;z) \;,
\]
\vspace{-0.5em}
is related to multiple polylogarithms \cite{Lappo}
\[
\Li{k_1,k_2, \ldots, k_n}{x_1, x_2, \ldots, x_n} = 
\sum_{m_n > m_{n-1} > \cdots> m_2 > m_1 > 0}^\infty \frac{x_1^{m_1}}{m_1^{k_1}} \frac{x_2^{m_2}}{m_2^{k_2}} 
\times\cdots\times \frac{x_n^{m_n}}{m_n^{k_n}}\;, 
\]
\vspace{-0.5em}
by
\vspace{-0.5em}
\begin{eqnarray}
G_{m_n, \ldots, m_1}\left(x_n, \ldots,  x_1; z \right)
&= &
(-1)^n 
\Li{m_1,m_2, \ldots, m_n}{\frac{x_2}{x_1}, \frac{x_3}{x_2}, \ldots, \frac{z}{x_n}}\;.
\nonumber \\  
\Li{k_1,\ldots, k_n}{y_1, \ldots, y_n} 
&= &
(-1)^n 
G_{k_n,\ldots, k_1}\left(\frac{1}{y_n}, \ldots, \frac{1}{y_1 \times\cdots\times y_n};1 \right)\;,
\nonumber 
\end{eqnarray}
where we have used
$ I(a_1, \ldots , a_k;z) = 
I\left(\frac{a_1}{z}, \ldots , \frac{a_k}{z};1\right) \;.  $

As an illustration of our technique, we shall consider the 
Appell hypergeometric function $F_1$ defined by the series
$ F_1(a, b_1, b_2, c; z_1,z_2) = 
\sum_{m=0}^\infty \sum_{n=0}^\infty
\frac{(a)_{m+n} (b_1)_m (b_2)_n}{(c)_{m+n}}
\frac{z_1^m}{m!} \frac{z_2^n}{n!} .  $
For this function, the differential reduction algorithm \cite{D}
is equivalent to the statement
that any function $F_1(a,b_1,b_2,c;z_1,z_2)$ can be expressed as 
a linear combination of function with arguments that differ from the 
original ones by an integer, 
$F_1(a\!+\!m_0;b_1\!+\!m_1;b_2\!+\!m_2;c\!+\!m_3;z_1,z_2)$, 
and its first derivatives
$ R_3 F_1(\vec{A}\!+\!\vec{m};z_1,z_2) = 
\left[ \sum_{j=1,2}  R_j \theta_j \!+\! R_0 \right]F_1(\vec{A};z_1,z_2)$, 
where $\theta_r = z_r \frac{d}{d z_r} \;,r=1,2$, 
$\vec{A} = (a,b_1,b_2,c)$ is a list of parameters, 
the $\vec{m}$ are lists of integers, and 
the $R_i$ are polynomials in parameters $a,b_r,c$ and $z_r$.
%We may rewrite the system of differential equation (\ref{diff:F1}) in terms 
of basis functions
%$\omega_0(z_1,z_2) = F_1(a\ep, b_1\ep, b_2\ep, 1\!+\!c\ep;z_1,z_2)$ 
%and differential operators $\theta_r = z_r \frac{d}{d z_r} \;,r=1,2$ as 
The basis function
$\omega_0(z_1,z_2) = F_1(a\ep, b_1\ep, b_2\ep, 1\!+\!c\ep;z_1,z_2)$ 
satisfies the system of differential equations
\begin{eqnarray}
\left[ 
(1\!-\!z_1) \frac{\partial}{\partial z_1}
\left( \theta_1 \!+\! \theta_2 \right) \right] \omega_0(z_1,z_2) & = &
\left[ \left( a\!+\!b_1 \!-\! \frac{c}{z_1} \right) \ep \theta_1
\!+\! b_1 \ep \theta_2 \!+\! ab_1 \ep^2 \right] \omega_0(z_1,z_2) 
\; ,
\label{diff:basis1}
 \\ 
\left[ 
(1\!-\!z_2) \frac{\partial}{\partial z_2} \left( \theta_1 \!+\! \theta_2 
\right) \right] \omega_0(z_1,z_2) & = & \left[ 
\left( a\!+\!b_2 \!-\! \frac{c}{z_2} \right) \ep \theta_2
\!+\! b_2 \ep \theta_1 \!+\! ab_2 \ep^2 \right] \omega_0(z_1,z_2) \;.
\label{diff:basis2}
\end{eqnarray}
Due to the analyticity of $F_1$
with respect to its parameters,
Eqs.~(\ref{diff:basis1}) -- (\ref{diff:basis2})
hold for every coefficient function $\omega_0^{(k)}(z_1,z_2)$ 
in the expansion
$\omega_0(z_1,z_2) = 1+\sum_{k=1}^\infty \omega_0^{(k)}(z_1,z_2) \ep^k$,
and the coefficient equations are
\begin{eqnarray}
(1 \!-\! z_1) \frac{\partial}{\partial z_1}
\left( \omega_1^{(j)}(z_1,z_2) \!+\! \omega_2^{(j)}(z_1,z_2) \right)
& = &
\left[ a \!+\! b_1 \!-\! \frac{c}{z_1} \right] \omega_1^{(j-1)}(z_1,z_2)
\!+\! b_1 \omega_2^{(j-1)}(z_1,z_2)  
\!+\! ab_1 \omega_0^{(j-2)}(z_1,z_2) \;, 
\label{system:1}
\\ 
(1 \!-\! z_2) \frac{\partial}{\partial z_2}
\left( \omega_1^{(j)}(z_1,z_2)  \!+\! \omega_2^{(j)}(z_1,z_2) \right)
& = &
\left[ a \!+\! b_2 \!-\! \frac{c}{z_2} \right] \omega_2^{(j-1)}(z_1,z_2)
\!+\! b_2 \omega_1^{(j-1)}(z_1,z_2) \!+\! ab_2  \omega_0^{(j-2)}(z_1,z_2) \;,
\label{system:2}
\\ 
z_r \frac{\partial}{\partial z_r} \omega_0^{(j)}(z_1,z_2) &=& 
\omega_r^{(j)}(z_1,z_2) \;, r=1,2\;, \label{system:3}
\end{eqnarray}
where we have introduce new functions
$ \omega_r(z_1,z_2)  =  \theta_r \omega_0(z_1,z_2)$, $ r=1,2$, 
which have $\ep$-expansions of the form
$\omega_r(z_1,z_2) = \sum_{k=1}^\infty \omega_r^{(k)}(z_1,z_2) \ep^k$, $r=1,2$.
To solve this system of first order differential equations, the boundary 
condition must be specified. Our choice is 
\[
\omega_r(z,z)=z\frac{ab_r\ep^2}{1+c\ep}{}_2F_1(1\!+\!a\ep,1\!
+\!(b_1+b_2)\ep;2\!+\!c\ep;z) \ \hbox{and}\ %
\omega_0(z,z) = {}_2F_1 (a\ep,(b_1\!+\!b_2)\ep;1\!+\!c\ep;z),\]
where the all-order $\ep$-expansion of the Gauss hypergeometric function
has the form \cite{diff_red,KWY07c} 
\[
{}_2F_1(1\!+\!A\ep,1\!+\!B\ep;2\!+\!C\ep;z) = \frac{1\!+\!C\ep}{z}
\sum_{j=0}^\infty \rho^{A,B,C}_j(z) \ep^j \;, \quad 
{}_2F_1(A\ep,B\ep;1\!+\!C\ep;z)    
= 1 \!+\! A B \ep^2 \sum_{j=0}^\infty W_j^{A,B,C}(z) 
\ep^j \;, 
\]
where $\rho^{A,B,C}_j(z)$ and $W_j^{a,b_1\!+\!b_2,c}(z)$ 
are expressible in terms of generalized polylogarithms. \cite{Lappo}
In particular, for the coefficient functions $\omega_r^{(j)}(z_1,z_2)$, we have
\[
\omega_1^{(j)}(z,z) \!+\! \omega_2^{(j)}(z,z) 
= a (b_1\!+\!b_2) \rho^{a,b_1+b_2,c}_{j-2}(z) \;, \quad 
\omega_0^{(j)}(z,z) =  a(b_1\!+\!b_2) W_{j-2}^{a,b_1\!+\!b_2,c}(z) \;.
\]
The solution of Eqs.~(\ref{system:1}),(\ref{system:2}) can now be written as
\[
\omega_1^{(j)}(z_1,z_2) \!+\! \omega_2^{(j)}(z_1,z_2) =
\frac{1}{2} a (b_1\!+\!b_2) \left[ \rho^{a,b_1\!+\!b_2,c}_{j-2}(z_1)  \!+\! \rho^{a,b_1\!+\!b_2,c}_{j-2}(z_2) \right] 
\!+\!  
\frac{1}{2} 
\int^{z_1}_{z_2} \frac{dt}{1\!-\!t} \left[ f^{(j-1)}(t,z_2) \!-\! 
h^{(j-1)}(z_1,t) \right] 
, 
\]
which should be supplemented by a self-consistency condition, 
\[
\int^{z_1}_{z_2} \frac{dt}{1-t} \left[ f^{(j-1)}(t,z_2) \!+\! 
h^{(j-1)}(z_1,t) \right] = 
a (b_1\!+\!b_2) \left[ \rho^{a,b_1\!+\!b_2,c}_{j-2}(z_1)  \!-\! 
\rho^{a,b_1+b_2,c}_{j-2}(z_2) \right] ,
\]
where 
$f^{(j-1)}(z_1,z_2)$ and $h^{(j-1)}(z_1,z_2)$ are symbolic notation for the 
r.h.s.\ of Eqs.~(\ref{system:1}), (\ref{system:2}), respectively.
The last equation (\ref{system:3}) may be rewritten as
$ \left( z_1 \frac{\partial }{\partial z_1 } \!+\! 
z_2 \frac{\partial }{\partial z_2 } \right) \omega_0^{(j)}(z_1,z_2) 
= \omega_1^{(j)}(z_1,z_2)  \!+\! \omega_2^{(j)}(z_1,z_2)$.
The solution may be obtained by the method of characteristics, and 
we obtain a first-order ordinary differential equation 
depending on parameter $C=\frac{z_2}{z_1}$: 
$ z_1 \frac{\partial }{\partial z_1 } \omega_0^{(j)} = 
\omega_1^{(j)}(z_1,C z_1)  \!+\! \omega_2^{(j)}(z_1,Cz_1)$.
Collecting all of these relations together provides an
iterative algorithm for constructing the 
all-order $\ep$-expansion of Appell hypergeometric function 
$F_1$ around integer values of the parameters. 
Due to the fact that all integral solutions contain only the factors 
$\frac{1}{1-t}$ and $\frac{1}{t}$,
we could expect that the result of iteration will always be 
expressible in terms 
of a  particular case of multiple polylogarithms.\cite{Lappo}

As an illustration, let us explicitly evaluate the first few coefficients. 
The first non-trivial term of the $\ep$-expansion corresponds to 
$j=2 (\omega_r^{(1)}= 0)$:
$ \omega_0^{(2)}(z_1,z_2) = \sum_{r=1,2} ab_r\, \Li{2}{z_r}$
and 
$ \omega_r^{(2)}(z_1,z_2) = - \sum_{r=1,2} a b_r \ln(1-z_r)$. 
The next iteration will produce
\begin{eqnarray}
\frac{1}{a} \omega_0^{(3)}(z_1,z_2)
& = & \sum_{r=1,2} b_r \left[ (a \!+\! b_r \!-\! c)\, \Snp{1,2}{z_r} 
- c\, \Li{3}{z_r} \right] 
+ b_1 b_2 \left[ \SN{1}{2}{z_1} \!+\! \SN{1}{2}{z_2} \!-\! 
\frac{1}{2} \ln^2 \left( \frac{1-z_1}{1-z_2} \right) \ln z_1  
\right]
\nonumber \\ && 
+ b_1 b_2 \Biggl\{ \left[ \Li{2}{\frac{z_2}{z_1} \frac{1\!-\!z_1}{1\!-\!z_2}} 
\!-\! \Li{2}{\frac{1\!-\!z_1}{1\!-\!z_2}} 
\right] \ln \left( \frac{1\!-\!z_1}{1\!-\!z_2} \right)
\!-\! \Li{3}{\frac{z_2}{z_1} \frac{1\!-\!z_1}{1\!-\!z_2}} 
\!+\! \Li{3}{\frac{1-z_1}{1-z_2}} \Biggr \} \;.
\end{eqnarray}

In summary, the hypergeometric function approach to Feynman diagrams provides
a promising route to organizing and calculating the higher-order processes 
which are becoming increasingly important in precision 
phenomenology. Conversely, mathematicians may find that results motivated by
high-energy physics will uncover new relationships among these classes of
functions, providing a fertile area for interaction between mathematics
and physics.
\vspace{-1em}

\begin{acknowledgments} 
\vspace{-1em}
S.A.\ Yost\ thanks the organizers of ICHEP08 for the opportunity to present this
poster, and acknowledges the hospitality and support of the $6^{\rm th}$ 
Simons Workshop in Mathematics and Physics at SUNY Stony Brook, where this
poster was partially prepared. 
M.Yu.\ Kalmykov's work was supported in part by BMBF Grant No.\ 05~HT6GUA.
B.F.L.\ Ward was partially supported by US DOE grant DE-FG02-05ER41399.
\end{acknowledgments}

\end{document}